\documentstyle[prl,aps]{revtex}
\newcommand{\beq}[1]{\begin{equation}\label{#1}}
\newcommand{\eeq}{\end{equation}}
\newcommand{\bear}[1]{\begin{eqnarray}\label{#1}}
\newcommand{\ear}{\end{eqnarray}}
\newcommand{\nn}{\nonumber}

\newcommand{\rf}[1]{(\ref{#1})}
\newcommand{\iso}{ {\cong } }

\newcommand{\N}{ \mbox{\rm I$\!$N} }

\def\C{\mbox{\rm {I\kern-.520em C}}}

\newcommand{\Spec}{ \mbox{\rm Spec} }

\newcommand\mustbe{\stackrel{!}{=}}

\begin{document}
\draft
\title{ On the fundamental length of quantum geometry
and the black hole entropy }
\author{
Martin Rainer\dag\P\footnote[1]{e-mail: mrainer@phys.psu.edu}
}

\address{\dag
Center for Gravitational Physics and Geometry,
104 Davey Laboratory, The Pennsylvania State University,
University Park, PA 16802-6300, USA}

\address{\P
Gravitationsprojekt, Mathematische Physik I,
Institut f\"ur Mathematik,Universit\"at Potsdam,
PF 601553, D-14415 Potsdam, Germany}

\date{March 21, 1999}

\maketitle
\begin{abstract}
The geometric operators of area, volume, and length,
depend on a fundamental length $\ell$ of quantum geometry
which is a priori arbitrary rather than equal to the
Planck length $\ell_P$.
The fundamental length $\ell$ and the Immirzi parameter $\gamma$
determine each other.
With any $\ell$ the entropy formula
is rendered most naturally in units
of the length gap $\sqrt{{\sqrt 3}/2}({\sqrt\gamma}\ell)$.

Independently of the choice of $\ell$,
the black hole entropy derived from quantum geometry
in the limit of classical geometry
is completely consistent with the Bekenstein-Hawking form.

The extremal limit of $1$-puncture states of the
quantum surface geometry corresponds rather to an extremal string
than to a classical horizon.
\end{abstract}
%
\hspace*{0.950cm} PACS number(s):\  04.60.-m \ 04.70.Dy

In \cite{ABCK} the black hole entropy for the quantum geometry of
the exterior horizon of a black hole has been calculated from
first principles. The area is formally proportional to the (real
modulus of) the Immirzi\cite{Im} parameter $\gamma$, parametrizing
certain affine transformations of the underlying phase space which
maintain the symplectic structure. The result of \cite{ABCK}
agrees for one particular value \beq{ent} \gamma=\gamma_0:={\ln
2\over \pi\sqrt 3} \eeq with the Bekenstein-Hawking (BH) formula.
In \cite{Kr} certain extremal states of the quantum horizon
geometry were examined. It was shown that, under the assumption
that the (semi-)classical limit of these states existed and
saturated a Bogomol'nyi type bound as for a classical Kerr black
hole, these states would yield $\gamma\neq\gamma_0$ providing an
apparent contradiction to the result of \cite{ABCK}. However, as
is pointed out below, these states do not admit a semiclassical
limit of quantum geometry. Hence one might think that $\gamma$
might scale and, in particular, take a different value for quantum
and classical geometry. But, apart from contradicting Occam's
razor, such a scaling of $\gamma$ would explicitly break the very
symplectic structure on which the quantization of geometry is
based. Indeed the results below show that the argument of
\cite{Kr} is independent of either $\gamma$ or the choice of the
fundamental length $\ell$. It only depends on a combination of the
two.

Hence here we will  argue that the result of \cite{ABCK} has to be
correct a fortiori, if and only if the state space is considered
in the classical limit of quantum geometry.

It should also be noted here that string theory calculations
reproduce the BH formula,
because like the BH they are performed in the same semiclassical
regime of quantum fields in a certain background geometry.

The classical limit of quantum geometry
could be characterized by discrete spectra of the
geometric operators like area, volume \cite{RoSm,DPRo}
and length \cite{Th}
becoming dense
on a sufficiently large state space.
({The area spectrum has been proven
rigorously to become dense in  \cite{AL1}.
It appears a reasonable conjecture that the volume
spectrum \cite{AL2} does so too, since both operators are quite analogous
in the sense that they are constructed as the vanishing regulator limit
of the square root of
appropriately regulated expressions quadratic and cubic
respectively in the momenta of the holonomy variables.
The length operator however
is quite different,
since it requires actually a tubular regularization
of the curve and the defining regulated expression involves
commutators of the holonomy variables and the volume operator.})

Quantum geometry is now just on the edge to make for the first
time contact with its classical limit, the example under
investigation being the spatial black hole horizon geometry and
the entropy of its state space. Entropy is distinguished by the
fact that it is physically dimensionless; it requires just
counting. Hence it is intrinsically a scale invariant entity. This
is of advantage when the classical limit is investigated. Once the
classical limit is understood completely, quantum geometry has
achieved the first fundamental derivation of the black hole
entropy from an underlying microscopic theory. However the
classical limit is quite tricky: If the state space of quantum
surface geometries is restricted to particular spin networks the
classical limit may fail to exist.

In the present approach we start with arbitrary fundamental
length $\ell$ and arbitrary $\gamma$.
After resuming the results
for entropy and area,
we try to understand better the origin of the distinguished value
obtained in \cite{ABCK}.

The exterior horizon $H$ of a classical black hole geometry is an
inner null boundary of its exterior space-time $M$. As in
\cite{ABCK} here we consider pullbacks of the phase space from $M$
to a spatial (say Cauchy-like) slice $\Sigma$ which hits $H$
transversally on $S:=\partial\Sigma\iso S^2$. Let us consider
canonical the pulled-back real phase space variables
$(^{\gamma}A,^{\gamma}E):=(\Gamma-\gamma K,\frac{1}{\gamma}E)$ on
$\Sigma$, where $\Gamma$ and $K$ are spin connection and curvature
1-forms and E is a 2-form dual to the triad on $\Sigma$. Let
$^{\gamma}\delta\equiv (\delta^{\gamma}A,\delta^{\gamma}E)$ denote
a tangent vector on phase space at $(^{\gamma}A,^{\gamma}E)$. Here
$^{\gamma}A$ is dimensionless and $^{\gamma}E$ has physical
dimension (length)$^2$. The symplectic structure is then given as
by
\bear{sympl}
\Omega|_{(^{\gamma}A,^{\gamma}E)}
\left ( ^{\gamma}\delta, ^{\gamma}\delta' \right )
&:=&
{1 \over \ell^2} \int_{\Sigma} {\rm Tr}
\left [ \delta^{\gamma}E\wedge\delta^{\gamma}A'
- \delta^{\gamma}E'\wedge\delta^{\gamma}A \right ]
\nn \\&&
- \frac{k}{2\pi} \int_{S}  {\rm Tr}
\left [ \delta^{\gamma}A\wedge\delta^{\gamma}A' \right ]
\ear
on the real phase, where $ k :={A_S\over \gamma \ell^2} $ and
$A_S$ is the area of $S$. It has been been shown in  \cite{ABCK}
how \rf{sympl} is related to an action with $U(1)$ boundary
Chern-Simons (CS) term representing a natural choice of boundary
conditions and such that the Einstein equations are reproduced
locally. Here the only difference to the symplectic structure used
in \cite{ABCK} is the normalization. Their normalization would
formally correspond to $\ell=8\pi\ell_P$ while in our approach
$\ell$ is just {\em some} microscopic fundamental length. (Note
that the classically convenient factor $8\pi$ of two times the
unit sphere area embedded in flat space for quantum geometry makes
no sense {\em a priori}.) It eventually becomes fixed only after
comparison with the semiclassical BH-entropy formula has been
made. Therefore, here the prefactor of the classical action before
quantization is left a priori arbitrary rather than fixing it bona
fide to the Newton coupling. As long as we do not compare the
microscopic quantum geometry to its classical limit in the
presence of some further (matter) fields, there appears to be no
necessity to fix it in any particular manner. If the quantum
theory of geometry is obtained by quantizing a classical action we
can not expect that the classical couplings are a priori the right
ones, while in fact they have to be chosen  {\em a posteriori}
such that they yield the correct values in the physical classical
limit. There is no obstruction that finally a classical limit of
quantum geometry recovers indeed all the mathematical structure of
the formal classical action put in before the quantization. But
the numerical values of the classical coupling constants are only
determined by experiments in the classical limit of geometry
rather than being predetermined by quantum geometry. The
asymptotic regime of observation within a classical geometry is
the only one which is directly accessible to us while the regime
of quantum geometry is physically quite different. Indeed even
string theory (while still in the realm of classical background
geometry) predicts already a different  coupling at the scale
$\ell_{s}$  where  residue dilatonic fields from extra dimensions
combine to couple to gravity. Therefore the microscopical coupling
constants of quantum geometry should only be determined in the
classical limit by consistency with the observed ones. With
$\ell<<\ell_{s}$, and $\ell_{s}$ close to $\ell_P$ within few
orders of magnitude only, it appears  rather implausible that
$\ell$ could equal $\ell_P$.

Consider now a finite set $P$ of discrete transversal punctures
(labeled $p=1,\ldots,N$) of $S$ by edges from a {\em gauge
invariant} spin network. The degrees of freedom on its surface are
initially given by a choice of a $SU(2)$-representation of spin
$j_p$ at each puncture $p$, induced by an intersecting edge of the
$SU(2)$ cylindrical state of the same spin $j_p$. Identifying the
transversal components of the $SU(2)$ connection in the
representation of spin $j$ along the intersecting edge with an
$U(1)$ connection, this then induces a $U(1)$ Chern-Simons state
on the surface, The trivial $U(1)$-representation has then a
multiplicity $2j+1$ which also is the dimension of the
corresponding Hilbert space. So on the $S^2$ surface the $SU(2)$
gauge symmetry is broken to $U(1)$. In the $SU(2)$ spin network
representation with spins $\{ j_p \}_{p\in P}$ at the punctures
$P$ on $S$, the area operator  is diagonal with eigenvalue
\beq{evarea} A^P_{S}:=\gamma\ell^2\sum_{p\in P} \sqrt{j_p(j_p+1)}
. \eeq Like in \cite{ABCK} counting of states fixes the entropy as
\beq{S} S^P={\gamma_0 \over \gamma \ell^2}{2\pi A^P_S } =\ln 2
{2\over\sqrt 3}  \sum_{p\in P}\sqrt{j_p(j_p+1)} , \eeq where the
$\gamma_0$ is given by \rf{ent}. Now in the classical limit the
result should agree with the BH formula. Hence in this limit
\beq{gamma} {\gamma} \ell^2 \mustbe 8\pi{\gamma_0} \ell^2_P  ,
\eeq which only fixes the combination ${\sqrt\gamma} \ell$.

The formula \rf{S} takes a more elegant form
when rendered in terms of
the length gap
\beq{lam}
\lambda:={\sqrt{{\sqrt 3}/2}} ({\sqrt\gamma}\ell) ,
\eeq
the lowest eigenvalue of the length operator.
In fact it would be assumed here
for any curve segment  in $S$
intersecting the divalent vertex
given by the puncture $p$ on $S^2$ of any edge
with spin $j_p=1/2$.
Let
\beq{alam}
a:= 2\pi \lambda^2
\eeq
be a half the area of of a sphere of radius $\lambda$.
Then the entropy formula simply reads
\beq{qgent}
e^{S^P}=2^{A^P_S/a} .
\eeq
So  \rf{lam} and \rf{alam}
render the microscopical entropy formula the
simple form \rf{qgent},
while \rf{gamma} guarantees
agreement with the BH formula
in the classical limit.
E.g. for a state with $N$ punctures of spin $\frac{1}{2}$
(a state from bits) it holds $A^P_S/a=4N$
and \rf{qgent} simply reads $e^{S^P}=2^{4N}$.

In order to illustrate the difficulty with the classical limit,
let us now consider the space $Q$ of surface states given by
a finite number of punctures
with all edges of same spin $j$ and restricted to meet $S$
transversally only.
The spectrum restricted to these states is
\beq{specNj}
\Spec:=\{\gamma\ell^2N\sqrt{j(j+1)} ,\ N,j\in \N_0 \} .
\eeq
Let us now consider the particular
subspace $Q_1$ given by the $1$-puncture states considered in \cite{Kr}.
Here $A_{S}:=\gamma\ell^2 \sqrt{j(j+1)}$ where $j$ is restricted
to integer values by gauge invariance since $S$ is the closed
inner boundary of $\Sigma$.
In the limit $j\to\infty$, the bound
\beq{qbound}
j < {A_S\over \gamma \ell^2} ,
\eeq
becomes saturated in leading order, i.e.
\beq{satlo}
{A_S \over \gamma \ell^2 j}-1 \to +0 .
\eeq
Under
the assumption
that $j\to\infty$ was a classical limit
for the surface geometry,
and
$J:=\hbar j$ corresponded here to a
classical angular momentum
one might postulate
that in the classical limit \rf{qbound}
corresponds to an asymptotic saturation of
\beq{cbound}
A_S /(8\pi G)>J/c^3 ,
\eeq
where $G$ is the Newton constant and
$c$ is the velocity of light.
(A bound like \rf{cbound} would e.g.
be satisfied for the spatial horizon area $A_S$
and angular momentum of a Kerr black hole.)
\rf{satlo} implied then in particular that
\beq{fixg}
\gamma=8\pi(\ell_P/\ell)^2 .
\eeq
Hence for {\em any} possible choice of $\ell$
there would be a contradiction between \rf{fixg}
and \rf{gamma} unless
$\gamma_0=1$ which then would seem to be in contradiction
to the calculated value of \cite{ABCK}.

However, the $1$-puncture configurations considered here are a
very particular subspace $Q_1$ of the full configuration space
only whence the entropy with respect to $Q_1$ has as pointed out
in \cite{Kr} indeed to be less than the entropy of \cite{ABCK}.

Moreover, these highly degenerate $Q_1$ configurations never yield
a classical limit for $j\to\infty$. The spectrum $\Spec_1$  of the
area operator on $Q_1$ has differences between neighboring
eigenvalues which asymptotically become $\gamma\ell^2$ for
$j\to\infty$. Hence ${\Spec_1}$ does not become dense in the limit
$j\to\infty$. (Note that this is unlike the length spectrum
\cite{Th}
 which becomes dense for $j\to\infty$ even when restricted
to $Q_1$ since the distance of neighboring eigenvalues there is
proportional to $1/\sqrt{j}$, for curve segments in $S$ containing
the puncture.)

In fact, since all the area is concentrated in a single point on
$S$, the limit $j\to\infty$ corresponds rather to an infinitely
extended string than to a horizon surface. The corresponding limit
state appears as the quantum analogue of a string hair leaving the
extremely degenerate horizon at the vortex. Geometrically, one
could indeed say that it is just a string hair without any
horizon. Abelian Nielsen-Olesen string hairs emerging from a
vortex were proven to exist and have been examined in quite some
detail within the Abelian Higgs model \cite{AGK}-\cite{BEG}. There
the $U(1)$ symmetry is broken everywhere but at the vortex by a
Higgs potential. For the surface states considered here the local
$U(1)$ gauge symmetry , which was obtained by particular boundary
conditions on $S$, on $Q_1$ spin networks is trivially also
global. However, the topological boundary $S=\partial\Sigma$ is
nowhere represented geometrically but at the puncture, where the
transverse intersecting edge plays the analogue of an axis of
rotation. Due to the asymptotic saturation of \rf{cbound}, it is
tempting to conjecture that the $j\to\infty$ limit of the quantum
string hair corresponds actually to a BPS state.

From above examples it is obvious that
the classical limit
of increasingly dense spectrum requires
to admit a variable number of punctures
and a mixture of different spins $j_p$
(this was used in the density proof
for the area operator given in \cite{AL1}).
Besides, at least in principle one has to take into account
also states with edges on the surface $S$
and trivalent punctures.

Once we are able to perform the classical limit
with a sufficiently large state space then
the counting of the number of surface states on $S$
will yield the entropy independent of $\gamma$
and independent of $\ell$.
When fitted to the BH formula in the classical limit,
the fundamental length $\gamma$ and the Immirzi parameter $\gamma$
are no longer independent since
in this limit the unknown
the interaction between quantum geometry and quantum matter
has to approach the known semiclassical coupling of
classical geometry to quantum matter a fortiori.
In fact $\gamma\ell^2$ enters the matter fields
and their momenta with a power determined by the
spin of the matter field\cite{GMM}.
One might hope that TQFT (see e.g. \cite{Ba})
may once be able to
provide a model for the quantization
of both geometry and matter. The former quantization
introduces necessarily the fundamental length $\ell$,
while the quantization of matter
has to yield usual field quantization which necessarily
invokes the Planck constant $\hbar$.
In fact it is only then that the Newton coupling
related to $G/c^3=\ell^2_P/\hbar$
becomes to play a nontrivial role.
However to fix the value of both $\ell$ and $\gamma$
requires something more, namely a more detailed
understanding {\em how} quantum geometry
converges towards its (semi-)classical limit.

Let us conclude emphasizing again, that at present all results
from quantum geometry are perfectly consistent with the BH formula
for the black entropy. With \rf{qgent} the entropy formula takes
also microscopically a simple form. While the classical limit
fixes the relation between $\gamma$ and the fundamental length
$\ell$, on the microscopic level quantum geometry and quantum
matter may still interact with a coupling different from the one
of the classical limit.
\section*{Acknowledgments}
I thank A. Ashtekar and K. Krasnov for useful comments,
and N. Guerras for communicating refs. \cite{AGK} -\cite{BEG}
to my attention.


\begin{thebibliography}{99}
%
\bibitem{ABCK}
A.\ Ashtekar, J.\ Baez, A.\ Corichi, K.\ Krasnov,
Quantum Geometry and Black Hole Entropy,
{\sl Phys.\ Rev.\ Lett.} {\bf 80}, No. 5, 904-907 (1998).
%
\bibitem{Im}
G.\ Immirzi,
Quantum Gravity and Regge Calculus,
{\sl Nucl.\ Phys.\ Proc.\ Suppl.} {\bf 57} 65-72 (1997).
%
\bibitem{Kr}
K. Krasnov,
Quanta of geometry and rotating black holes,
gr-qc/9902015.
%
\bibitem{RoSm}
C.\ Rovelli and L.\ Smolin,
Discreteness of area
and volume in quantum gravity, {\sl Nucl.\ Phys.} {\bf B442},
593 (1995); Erratum: {\sl Nucl.\ Phys.} {\bf B456}, 734 (1995).
%
\bibitem{DPRo}
R.\ De Pietri and C.\ Rovelli,
Geometry Eigenvalues and Scalar Product
from Recoupling Theory in Loop Quantum Gravity, {\sl Phys.\ Rev.} {\bf D54},
2664 (1996).
%
\bibitem{Th}
T.\ Thiemann,
A length operator for canonical
quantum gravity,  gr-qc/9606092.
%
\bibitem{AL1}
A.\ Ashtekar and J.\ Lewandowski, Quantum Theory of Geometry I: Area
operators, {\sl Class.\ Quant.\ Grav.} {\bf 14}, 55 (1997).
%
\bibitem{AL2}
A.\ Ashtekar and J.\ Lewandowski, Quantum Theory of Geometry II:
Volume Operators, {\sl Adv.\ Theor.\ Math.\ Phys.} {\bf 1}, No. 2 (1998).
%
\bibitem{AGK}
A. Achu\u carro, R. Gregory, and K. Kuijken,
Phys. Rev. D {\bf 52}, 5729-5742 (1995).
%
\bibitem{AG}
A. Achu\u carro and R. Gregory,
{\sl Phys.\ Rev.\ Lett.} {\bf 79},  1972-1975 (1997).
%
\bibitem{CAES}
A. Chamblin, J.M.A. Ashbourn-Chamberlin, R. Emparan, and A. Sornborger,
Phys. Rev. D {\bf 58}, 124014 (1998).
%
\bibitem{BEG}
F. Bonjour, R. Emparan, R. Gregory,
Vortices and extreme black holes: the question of flux expulsion,
gr-qc/9810061.
%
\bibitem{GMM}
L.J. Garay, G.A. Mena Marugan, Class.Quant.Grav. 15 (1998) 3763-3775
%
\bibitem{Ba}
J. Baez,
Higher-Dimensional Algebra and Planck-Scale Physics,
gr-qc/9902017.


\end{thebibliography}
\end{document}